\documentclass[article]{aastex61}
\usepackage{hyperref}
\usepackage{ textcomp }
\usepackage[]{natbib}

\newcommand{\kms}{{km~s$^{-1}$}}

\newcommand{\moy}{$M_\odot$~yr$^{-1}$}

\shorttitle{Novel Mass Loss Measurements}
\shortauthors{Kobulnicky et al.}

\begin{document}

\title{DEMONSTRATION OF A NOVEL METHOD FOR MEASURING MASS-LOSS RATES FOR MASSIVE STARS}

\correspondingauthor{Henry A. Kobulnicky}
\email{chipk@uwyo.edu}
\author{Henry A. Kobulnicky}
\author{William T. Chick}
\affiliation{Department of Physics \& Astronomy, University of Wyoming, 
  Dept 3905, Laramie, WY 82070-1000, USA}
\author{Matthew S. Povich}
\affiliation{Department of Physics \& Astronomy, California State 
 Polytechnic University, 3801 West Temple Avenue,  Pomona, CA 91768, USA}

\begin{abstract} 

The rate at which massive stars eject mass in stellar winds
significantly influences their evolutionary path.  Cosmic
rates of nucleosynthesis, explosive stellar phenomena, and
compact object genesis depend on this poorly known facet of
stellar evolution. We employ an unexploited 
observational technique for measuring the mass-loss rates of
O- and early-B stars. Our approach, which has no adjustable
parameters, uses the principle of pressure equilibrium
between the stellar wind and the ambient interstellar medium
for a high-velocity star  generating an infrared  bowshock
nebula.  Results for twenty bowshock-generating stars show
good agreement with two sets of theoretical predictions for
O5--O9.5 main-sequence stars, yielding $\dot
M=$1.3$\times$10$^{-6}$ to 2$\times$10$^{-9}$ \moy.  
Although $\dot M$ values derived for this sample are smaller
than  theoretical expectations by a factor of about two,
this discrepancy is greatly reduced compared to canonical
mass-loss methods.   Bowshock-derived mass-loss rates are
factors of ten smaller than H$\alpha$-based measurements
(uncorrected for clumping) for similar stellar types and are
nearly an order of magnitude larger than P$^{4+}$ and some
other UV absorption-line-based diagnostics. Ambient
interstellar densities of at least several cm$^{-3}$ appear
to be required for formation of a prominent infrared
bowshock nebula.   $\dot M$ measurements  for early-B stars
are not yet compelling owing to the small number in our
sample and the lack of clear theoretical predictions in the
regime of lower stellar luminosities.  These results may
constitue a partial resolution of the extant ``weak-wind
problem'' for late-O stars. The technique shows promise for
determining mass-loss rates in the  weak-wind regime.     

\end{abstract}


\section{Introduction} \label{sec:intro}

Massive stars shed mass prodigiously via their
radiation-driven stellar winds \citep{Lucy1970, Castor1975,
Pauldrach1986} and perhaps even more dramatically through
pulsationally driven  ejection events
\citep{Glatzel1999,Kraus2015,Yadav2017}. For single massive
stars, the mass-loss rate, $\dot M$, integrated over the
star's lifetime  determines its final act (i.e., type of
supernova or $\gamma$-ray burst),   end product (i.e.,
neutron star, black hole, none),  and its radiant and
nucleosynthetic contribution to the cosmos.  For the
$\simeq$50\% of massive stars having a close companion
\citep{Sana2012,Kobulnicky2014}, the  effects of mass
exchange and common envelope evolution are expected to be a
more significant evolutionary influence, but wind-driven
mass loss must still play a large role when integrated over
the lifetime of a star.  Observational and theoretical work
broadly agree that mass-loss rates range from $<10^{-8}$
\moy\ for weak-winded late-O stars to few$\times$10$^{-5}$
\moy\  for the most luminous evolved massive stars, with
rates being factors of several lower at low metallicities
\citep{Vink2001,Martins2005b,Fullerton2006,Mokiem2007,Muijres2012,Massa2017}.  
Luminous blue variables and related objects in unstable
phases of evolution may eject shells  from several tenths to
several solar masses in discrete eruptive events, resulting
in time-averaged rates of 0.01--few \moy\  
\citep{SmithOwocki2006}.  Such large excretion events 
influence not only the  evolution of the star but the
appearance of subsequent explosive phenomena, such as when
fast supernova ejecta encounter dense circumstellar
material, creating unusually luminous supernovae
\citep{Smith2007,Miller2009,Chevalier2011}. Given that most
massive stars are also members of (close!) multiple systems
\citep{Kobulnicky2007,Sana2012,Kobulnicky2016}, companion
interactions are certain to play an important (but poorly
characterized, at present) role. Reviews of massive star
winds and mass loss include \citet{Kudritzki2000},
\citet{Puls2008}, and \citet{Smith2014}.

Mass-loss rates for massive stars have been measured using a
variety  of techniques.  These include observations of
H$\alpha$ recombination lines
\citep{Leitherer1988,Lamers1993,Puls1996,Markova2004,Martins2005b},
radio-continuum and infrared free-free emission 
\citep{Abbott1981,Nugis1998,Puls2006,Massa2017}---so-called 
$n^2$ diagnostics because  the excess flux scales as the
square of the density of material in the wind.   Massive
star winds are demonstrably not ideal isotropic structures
with smooth density gradients.  Observational signatures of
density inhomogeneities (i.e., ``clumping'')  in OB star
winds  are abundant, including time-variable line profiles
and discrete absorption components
\citep{Ebbets1982,Fullerton1996,Lepine1999,Prinja2002}, 
especially in supergiants, and the presence of large-scale
optically-thick clumps \citep{Prinja2010}.  Accordingly, the
presence of clumps, which result inevitably from
instabilities in the line-driven wind  
\citep{Owocki1988,Dessart2005,Muijres2011}, may skew observationally determined mass
loss estimates upward relative to unclumped calculations. 
Although theoretical models for stellar winds include
provisions for a clumping in a heuristic way
\citep[e.g.,][]{Hillier1998,Puls2005},  the poorly
constrained clump geometries and kinematics introduce
uncertainties of one or two orders of magnitude in the
mass-loss rates from $n^2$ diagnostics.  Furthermore, the
traditional $n^2$ diagnostics become insensitive to mass
loss below rates of about 10$^{-7}$ \moy\
\citep{Markova2004,Mokiem2007,Marcolino2009}, corresponding
to luminosities of about  10$^{5.2}$ (approximately an
O7.5V), demanding other diagnostics for the weaker winds
expected from  less luminous stars.  

Ultraviolet spectroscopy of metal resonance lines  such as
\ion{C}{3}, \ion{N}{5}, \ion{Si}{4}, and \ion{P}{5} has
provided the other major diagnostic of mass-loss rates
\citep{Garmany1981,Howarth1989,Fullerton2006,Marcolino2009}.
UV-based estimates of mass-loss are less sensitive to 
clumping because they depend linearly on density as long as
the optically depths in the line cores are small and the
dominant ionization species can be observed.  Mass-loss
rates derived from  UV resonance lines are often factors
several to hundreds lower than $n^2$ diagnostics
\citep{Fullerton2006} in the limited  regimes where the two
methods overlap.  For mass-loss rates greater than  about
10$^{-7}$ \moy\ resonance lines begin to become optically
thick and derived $\dot M$ values become less certain,
especially if clumping is optically thick
\citep{Prinja2010}.  UV estimates may also be systematically
low   if coronal X-rays produce additional photoioinization
of  the metal ionic species probed by UV spectra and ionization
correction factors are not properly applied    
\citep{Waldron1984,Marcolino2009,Huenemoerder2012}. 

Theoretical mass-loss rates predicted on the basis of the
``modified wind momentum''\footnote{The modified wind
momentum is $\dot M v_{\infty} (R_*/R_\odot)^{0.5}$, where
$v_\infty$ is the terminal stellar wind  velocity, and $R_*$
is the stellar radius, after \citet{Puls1996}. }
\citep{Puls1996,Vink2001} agree with observations in the
limit of  strong winds and luminous stars (i.e.,
$\log(L/L_\odot)\gtrsim5.2$, spectral type earlier than
about O7V), but in the limit of weak winds ($\dot
M\lesssim10^{-8}$ \moy, $\log(L/L_\odot)\lesssim5.2$),
UV-derived mass-loss rates are lower than theoretical
predictions by up to two orders of magnitude---a discrepancy
known as the ``weak wind problem'', discussed extensively in
the literature 
\citep{Martins2005b,Mokiem2007,Marcolino2009,Muijres2012}.
Whether the discrepancy results from the effects of
clumping, unexpected ionization structure, variations in
$\dot M$ as a star evolves, limitations in the theoretical
treatment of the wind \citep{Lucy2010a,Krticka2017}, or some
combination, remains a matter of debate.   In the limit of
very massive and luminous supergiants near 50 M$_\odot$ the
\citet{Vink2012} ``transition mass-loss rate'' near
10$^{-5}$ \moy\ suggests that the current reductions of 2--3
in model mass-loss rates is appropriate.  However, much of
the O-star regime remains uncertain.  The recognition  that
some late-O stars exhibit much weaker winds than other O
stars of the same spectral type  is regarded as a kind of
second-order weak wind problem \citep{Marcolino2009} that
might be solved  along with with the resolution of the
canonical weak wind problem. \citet{Huenemoerder2012}
presented one possible resolution in their study of the
weak-wind O9.5V runaway star $\mu$ Col which evinces a
massive hot stellar wind visible in X-rays but only
tenuously detectable using UV metal absorption-line
spectroscopy.

Given the lingering order-of-magnitude uncertainties on 
mass-loss rates, together with the
sensitivity\footnote{Changing mass-loss rates by factors of
two or less can dramatically alter the sequence of stellar
evolutionary phases,  final masses,  stellar endpoints, and
nucleosynthetic yields
\citep[e.g.,][]{Meynet1994,Renzo2017,Meynet2015}!}  of
stellar and cosmic evolution to these values, alternative
observational diagnostics for $\dot M$ are warranted.
\citet{Kobulnicky2010} proposed using runaway
\citep{Blaauw1961,Gies1986} massive stars and the their
interstellar bowshock nebulae 
\citep{vanBuren1988,Noriega1997,Gvaramadze2008} as a new
laboratory for measuring mass-loss rates.  Following the
reasoning first articulated by \citet{Gull1979} for the
prototypical bowshock runaway star $\zeta$ Oph, we
employ the principle of balancing  the momentum flux
between the stellar wind and the impinging interstellar
material,

\begin{equation} 
 \rho_{w} V_w^2 =  \rho_{a} V_a^2 .
\end{equation}

\noindent Here, $\rho_w$ is the density of the stellar wind,
$V_w$  is the velocity of the wind, $\rho_a$ is the ambient
interstellar density, and $V_a$ is the velocity of the
ambient ISM in the rest frame of the star.   We make the
assumption that the stellar wind is isotropic, \citep[but
mass loss could be enhanced along the polar axis or reduced
at the equatorial plane for rapidly rotating stars,
][]{Owocki1996,Langer1998,Mueller2014}  so that the density
of the stellar wind can be expressed as,

\begin{equation}
\rho_w = {\dot{M} \over {4\pi R_0^2 V_w}} ~ ~ ,
\end{equation}

\noindent where $\dot{M}$ is the stellar mass-loss rate and
$R_0$ is the  ``standoff'' radius---the distance between the
star and the point where the momentum fluxes are equal. By
substitution of Equation 2 into Equation 1 and rearranging, the mass loss
rate can be expressed in terms of observable stellar and
interstellar properties,

\begin{equation}
 \dot M = { {4\pi R_0^2 V_a^2 \rho_a} \over {V_w}} .
\end{equation}

\noindent  $R_0$ is simply $R_{0r} D$, the angular size of
the standoff distance in radians times the distance to the
star.  The former is straightforward to measure from
infrared images, modulo the unknown factor for inclination;
arguably $\sin i\simeq 1$  in order that the bowshocks be
identified as arcuate nebulae \citep{Kobulnicky2017}. 
Distances, $D$, may be obtained by spectroscopic parallax,
or by geometric parallax measurements
\citep{Perryman1997,Gaia2016a,Gaia2016b}.  $V_w$  is taken
to be the terminal stellar wind speed, $V_\infty$,
appropriate to the spectral type and luminosity class, as
tabulated in the literature \citep[e.g.,][]{Mokiem2007},
although individual stars may  vary significantly about the
mean.   $V_a$ is expected to average 30 \kms\ for runaway
stars, but individual values may again vary significantly. 
``In-situ'' bowshocks \citep{Povich2008,Sexton2015} where the relative motion is caused by
an outflow of interstellar material at 10--15 \kms\
from an \ion{H}{2} region \citet[e.g., the Carina
star-forming region,][]{Walborn2002}  rather than a runaway
star, could be such exceptions.  Specific space velocities
for any star may ordinarily be calculated from measurements
of proper motions, radial velocities, and  distances.   The
ambient ISM density is the most challenging quantity to
measure.  \citet{Gvaramadze2012} applied this technique,
using the size of the \ion{H}{2} region surrounding $\zeta$
Oph and its ionizing flux to eliminate the ambient density
in Equation 3 (or equivalently, solve for it numerically).
They derived a mass loss  rate 2.2$\times$10$^{-8}$ \moy,
comparable to updated theoretical predictions of
\citet{Lucy2010a}.  However, they note that  \ion{H}{2}
regions are not generally present around bowshock-producing
stars, limiting the utility of this approach to measuring
$\rho_a$.  Here, we use the peak infrared surface brightness
of the nebula to estimate $\rho_a$, as described in
subsequent sections.  

In this contribution we apply the principle of momentum
balance to derive mass-loss rates for the 20
bowshock-producing stars having well-characterized stellar
parameters from Table~5 of \citet{Kobulnicky2017}.   As an
independent technique for estimating mass loss, our method
does not depend on requirements like optically thin atomic
lines, the adopted geometrical parameterization of clumping,
or a detailed treatment of the ionization structure in the
wind.  Massive stars are known to exhibit temporal
variability in their line profiles, indicating probable
variation in clumping and mass loss. We expect that our
approach has the added benefit of averaging over  short-term
fluctuations in wind structure and mass-loss rate.   This
method will undoubtedly  entail a different suite
uncertainties and potential biases than the traditional
ones, including the difficult-to-parameterize effects a 
star moving through non-uniform ambient ISM, a possibility
that we neglect in this initial treatment.  
Nevertheless, most of our targets are late-O dwarf stars,
making this sample especially relevant for addressing the
weak-wind problem.  In Section~2 we describe our methods for
determining the requisite  stellar and interstellar
parameters.  In Section~3 we  present new mass-loss rates
for this well-characterized sample of stars and compare them
to existing  observational and theoretical determinations
for stars of similar luminosity and evolutionary stage.
Section~4 summarizes implications for  these new results and
outlines prospects for future progress.

\section{Measuring $\dot M$}  
\subsection{Sample Selection}  

\begin{figure}
\plotone{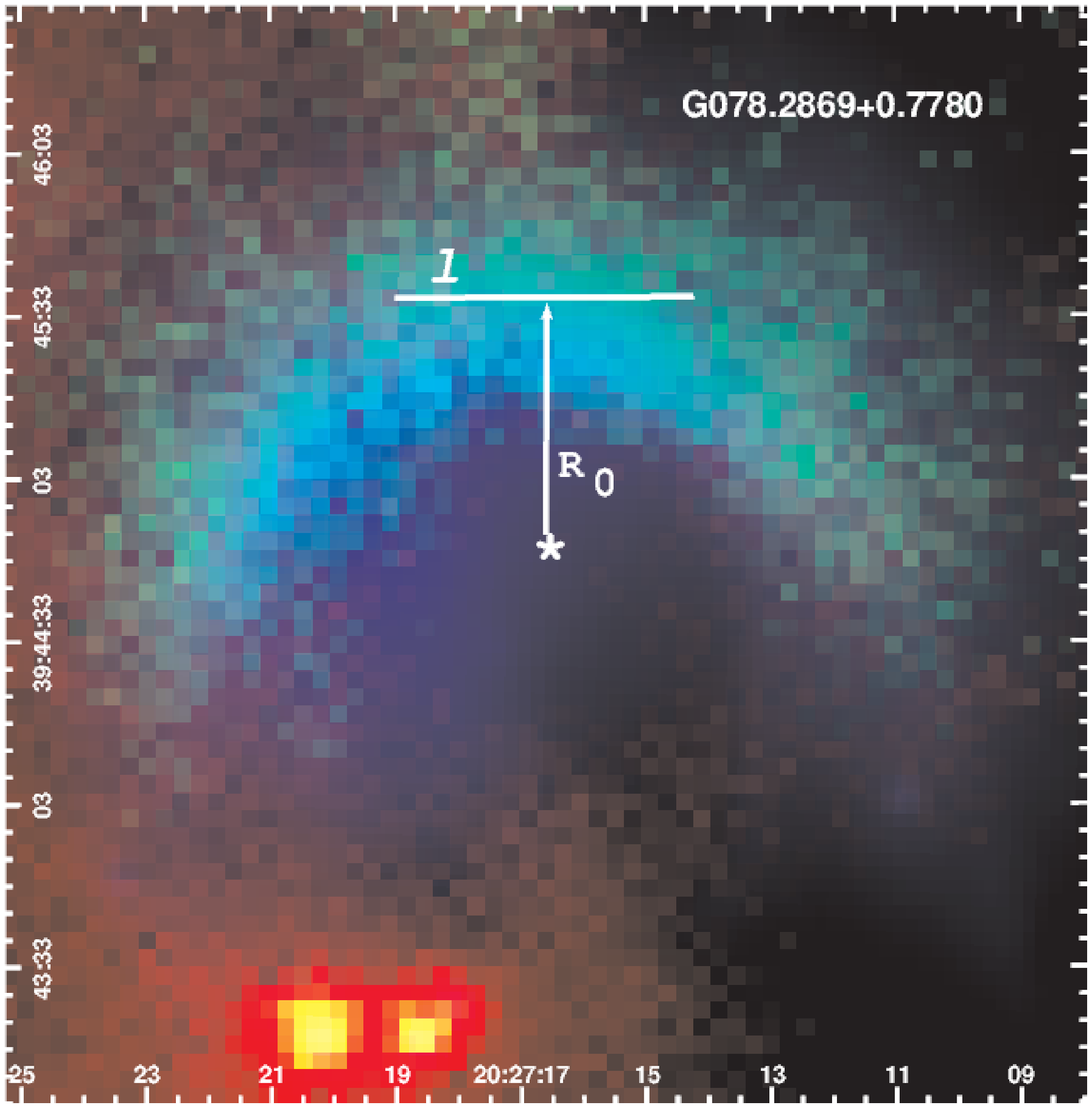}
\caption{Three-color representation of the G078.2869+00.7780 nebula (object \#331
from the catalog of \citet{Kobulnicky2016}) with blue/green/red representing 24/70/160 $\mu$m
from $SST$/$HSO$/$HSO$, respectively.   
White lines represent the standoff distance, $R_0$, and the chord
diameter, $\ell$. 
\label{fig:ExampleFig}}
\end{figure}

Table~1 lists the 20 stars from Table~5 of
\citet{Kobulnicky2017} selected from among the 709 bowshock
candidates of \citet{Kobulnicky2016} as having secure
distances, spectral types, and infrared photometric
measurements\footnote{Photometric data include a measurement
at the {\it Wide-Field Infrared Explorer (WISE)} 22 $\mu$m 
band or {\it Spitzer Space Telescope (SST)} 24 $\mu$m band
and the {\it Herschel Space Observatory (HSO)} 70 $\mu$m
band.} at multiple bandpasses covering the adjoining
nebula.  Column 1 contains the index number using the
numeration of \citet{Kobulnicky2016}. Column 2 lists common
name of the star, followed by the generic name in galactic
coordinates in column 3.  Columns 4 and 5 provide the
adopted spectral type/luminosity class and corresponding
literature references, respectively.  Only 3--4 objects have
evidence for being part of a multiple-star system.  This is
significant because it means that, in the majority systems,
one star is the dominant source of stellar wind, thereby
simplifying any the ensuing interpretation.  Columns 6 and 7
contain the adopted effective temperatures and radii, using
the theoretical O-star temperature scale (Tables 1---3) of
\citet{Martins2005a}.   For the few B stars we use the
temperatures and radii of \citet{Pecaut2013}.   Column 8
gives the adopted stellar mass. Column 9 provides the
adopted terminal wind speed calcualted by averaging galactic
O and B stars of the same spectral type from Table A.1 of
\citet{Mokiem2007} and Table 3 of  \citet{Marcolino2009}.
These values are uncertain at the level of 50\%, based on
the dispersion among multiple measurements at a given
spectral type.  Early-B dwarfs are particularly uncertain
owing to the difficulty in measuring wind lines. Column 10
lists the adopted distance along with the corresponding
reference in column 11. Most the sources have  distances 
estimated through their association with a star cluster or
molecular cloud having distance measurements from eclipsing
binaries (such as in Cygnus OB2),  main-sequence fitting, or
radio VLBI geometric parallaxes of masers within the
adjoining star-forming complex.  The O8V KGK2010 10 deserves
special mention.  At $\ell=$77\fdg0, $b$=-0\fdg6, it lies
more than two degrees from the main body of the Cygnus OB2
Association---outside the nominal boundaries of the Cygnus-X
star-forming complex, and possibly at a different distance. 
At $V=15$ mag it is also 2--3 magnitudes fainter at V band
than other O8V stars of similar reddening
\citep[c.f.,][Table 5]{MT91} in Cygnus OB2.  It may be a
background object at a distance similar to Cygnus X-3
\citep[][3.4 kpc or 9.3 kpc]{Ling2009}.  Based on its
similar reddening ($A_{\rm V}=$5.4 mag) to other
V$\approx$12.5 O8V stars in Cygnus OB2 at 1.32 kpc
\citep{Kiminki2015}, we use the factor of 9.8 in flux,
corresponding to a factor of 3.1 in distance, to adopt a
scaled distance of 4.1 kpc, while cautioning  of a more
uncertain distance for this object.  Column 12 lists the
standoff distance, $R_0$, in arcsec, while column 13 lists
the standoff distance in pc, calculated from  the distance
in column 10 and the angular separation.\footnote{We apply a
statistical correction factor of  1/sin(65\degr)$=$1.10 for
geometrical projection effects when computing $R_0$ in pc.
This is a suitable correction because  inclinations
substantially smaller than about 50\degr\ would begin to
mask the  bowshock morphology and make the object unlikely
to be included in the list of bowshock candidates (e.g., see
\citet{Acreman2016} for numerical simulations of bowshocks
at various inclination angles.)}  Column 14 lists the peak
$HSO$ 70 $\mu$m surface brightness above adjacent background
levels in Jy sr$^{-1}$, occurring  at a location near the
apex of the nebula. We use the 70 $\mu$m measurement because
the majority of bowshock candidates are detected and
stochastic heating effects are unlikly to affect dust at
this wavelength compared to 24 $\mu$m.    Column 15 lists
angular  diameter in arcsec of the nebulae along a chord
($l$) intersecting the peak surface brightness.
Figure~\ref{fig:ExampleFig} shows the G078.2889+00.7829
nebula surrounding the central star LSII+39~53 (O7V). 
Red/green/blue depict the $HSO$ 160 and 70 $\mu$m and $SST$
24 $\mu$m data, respectively. The white asterisk marks the
location of the central star.   White lines represent the
standoff distance, $R_0$, and the chord diameter, $\ell$.

\subsection{Calculation of  $\dot M$}  

Beginning with Equation~3, we express the standoff distance,
$R_0$ as the product of the distance to the source, $D$,
times the angular distance from star to apex in radians,
$R_{0r}$.   Given that very few bowshock stars  have
measured space velocities, we adopt for $V_a$ a typical
``runaway'' speed  of 30 \kms\ \citep{Gies1986}. For $\zeta$
Oph we adopt the 26.5 \kms\ calculated from its proper
motion and radial velocity \citep{Gvaramadze2012}.
Velocities much smaller than this are unlikely to  produce
bowshocks while stars moving much faster than this would
fall on the tail of the Maxwell-Boltzmann velocity
distribution and be  quite rare.   Thirty kilometers per
second agrees well with the measured space velocity for the
nearest bowshock star, $\zeta$ Oph \citep{Gvaramadze2012},
and is  a reasonable average value in the absence of
individual data.   The ambient interstellar density,
$\rho_a$, can be expressed as  $n_a\bar m$, the ambient
number density and mean particle mass, respectively.   We
adopt $\bar m=2.3\times10^{-24}$ g, appropriate to the Milky
Way interstellar medium (ISM).  While the ambient number
density $n_a$ preceding the bowshock is challenging to
measure, the density within a bowshock nebula, $n_N$, can be
estimated from knowledge of the infrared surface brightness
(i.e., specific intensity), the dust emissivity, the
line-of-sight path length through the nebula, and a
reasonable assumption for the gas-to-dust ratio.   $I_\nu$
is the specific intensity at a selected infrared frequency
in erg s$^{-1}$ cm$^{-2}$ sr$^{-1}$,  such as the $HSO$ 70
$\mu$m band, where the optical depth of dust is thought to
be low for essentially all Galactic sightlines.  $I_\nu$ is
directly measurable from the $HSO$ images in Jy sr$^{-1}$.  
The path length through the dust nebula is $\ell$,
calculated from the projected angular  diameter of the chord
in radians, $\ell_{r}$,  times the adopted distance, $D$. 
This assumes rotational symmetry of the nebula such that the
line-of-sight depth is adequately represented by the
projected diameter.   The dust emission coefficient  per
nucleon, $j_\nu$ in Jy cm$^2$ sr$^{-1}$ nucleon$^{-1}$ as
tabulated by \citet{DL07}\footnote{Note, \citet{DL07} use
the emission coefficient, $j_\nu$, to express the  energy
emitted per second per steradian per Hertz per {\it nucleon}
instead of the  conventional definition of energy
emitted per second per steradian per Hertz {\it per volume}.},
is determined by fitting their dust models to the infrared
spectral energy distribution for each object, as performed
in \citet{Kobulnicky2017} for the tabulated list of 20
objects.  The \citet{DL07}  Milky Way dust models assume a
standard 1:136 dust grain to atomic particle (principally H
+ He) ratio, so that the models provide directly the desired
nucleon number density within the infrared nebulae,

\begin{equation}
n_N  = { I_\nu \over {\ell ~j_\nu}  }.
\end{equation}

\noindent  This is really an {\it average} density along
the  chord length intersecting the bowshock apex; the peak
density near the apex of the bowshock could be larger.   
However, we desire to measure the {\it ambient} ISM density
preceding the bowshock, $n_a$, not $n_N$.  A physically
motivated conversion  would be $n_a=0.25 n_N$, given that
the density increases by a factor of four across a strong
shock \citep[e.g.,][]{Landau1987}, which we expect given 
the highly supersonic nature of the $>$1000 \kms\ stellar
winds. Therefore,

\begin{equation}
n_a  = {{0.25 I_\nu} \over {\ell ~j_\nu}  }.
\end{equation}

With these substitutions
Equation 3 then becomes 

\begin{equation}
 \dot M = { {4\pi {R^2_{0r}} D^2 V_a^2 I_\nu 0.25 {\bar m}} \over {V_w \ell_r D j_\nu}} ,
\end{equation}
 
\noindent which eliminates once factor of distance simplifies to

\begin{equation}
 \dot M = { {\pi {\bar m} {R^2_{0r}} D V_a^2 I_\nu} \over {V_w \ell_{r}  j_\nu}} .
\end{equation}
 
\noindent In convenient astrophysical units this can be expressed as a mass-loss rate in solar masses per year,

\begin{equation}
 \dot M (M_\odot ~yr^{-1}) = 1.67\times10^{-28} { {   {[R_{0}(arcsec)]^2} D(kpc)~[V_a(km~s^{-1})]^2 ~I_\nu(Jy~sr^{-1})} 
  \over {V_w(km~s^{-1}) ~\ell(arcsec) ~ j_\nu(Jy~cm^2~sr^{-1} nucleon^{-1})}} .
\end{equation}

\noindent This expression is linear in all of the crucial
quantities except the relative velocity between star and ISM
and the angular standoff distance.  The angular quantities
$R_{0r}$ and $\ell_{r}$ are measured to about 10\% from
infrared images.  $I_\nu$ is measured to about 20\%, the
dominant source of uncertainty being the definition of a
suitable local background level.  Mean stellar wind
velocities are also measured to, perhaps, 50\%.  For this
sub sample distances are known to better than 25\% in the
majority of cases.  The stellar space velocities are 
approximations based on mean values for runaway stars;
these may not be known to better than a factor of two.  
This makes the error budget on  $\dot M$ nearly 60\%, neglecting
stellar velocity uncertainty, or about factor of two with it
included.   The model-dependent choice of $j_\nu$ is the
most significant remaining variable.  

\cite{DL07} provide a grid of models for interstellar dust,
yielding $j_\nu$ as a function of wavelength and as a
function of  incident radiant energy density, $U$.  The
models are parameterized  in terms three variables: 
$U_{min}$, the minimum radiant energy density\footnote{$U$
is defined in \citet{DL07}  as the ratio of the incident
radiant energy density (in erg~cm$^{-3}$) to the mean
interstellar radiant energy density estimated by
\citet{Mathis1983}.} to which dust is exposed, a maximum
radiant energy density $U_{max}$ to which dust is exposed,
and a fraction of PAH molecules by mass, $q_{PAH}$, within
the material.  \citet{Kobulnicky2017} fitted DL07 models to
the infrared spectral energy distributions of objects from
Table~1, concluding that, in the majority of cases, models
with a single radiant  energy density (i.e., 
$U_{min}=U_{max}$) provided the best match to the data. 
They also tabulated an estimate for $U$ in each nebula,
calculated from the star's luminosity and standoff
distance, assuming that the central star was the dominant
source of  illumination.  Typical values for $U$ ranged from
few$\times$10$^2$ to few $\times$10$^4$, lending credence to
the proposition that the central star  dominates the
energetics of each nebulae.      Furthermore, models with
the minimum PAH mass fraction, $q_{PAH}$=0.47\%,  were
preferred, suggesting that PAHs are either
destroyed or not present in the bowshock nebulae. 
Accordingly, we adopt for each object the emission
coefficient, $j_\nu$, given for the bandpass-averaged  $HSO$
$PACS$\footnote{Photoconductor Array Camera and Spectrometer
\citep[PACS;][]{Poglitsch2010}.}  70 $\mu$m band from the
single-$U$ DL07 model appropriate to the  $U$ for each
object.   Because DL07 provide models only at discrete 
radiant energy densities of $U$=10$^2$, 10$^3$, 10$^4$, 10$^5$,
we employ linear interpolation to obtain a $j_\nu$
appropriate to each object.  

\begin{figure}
\plotone{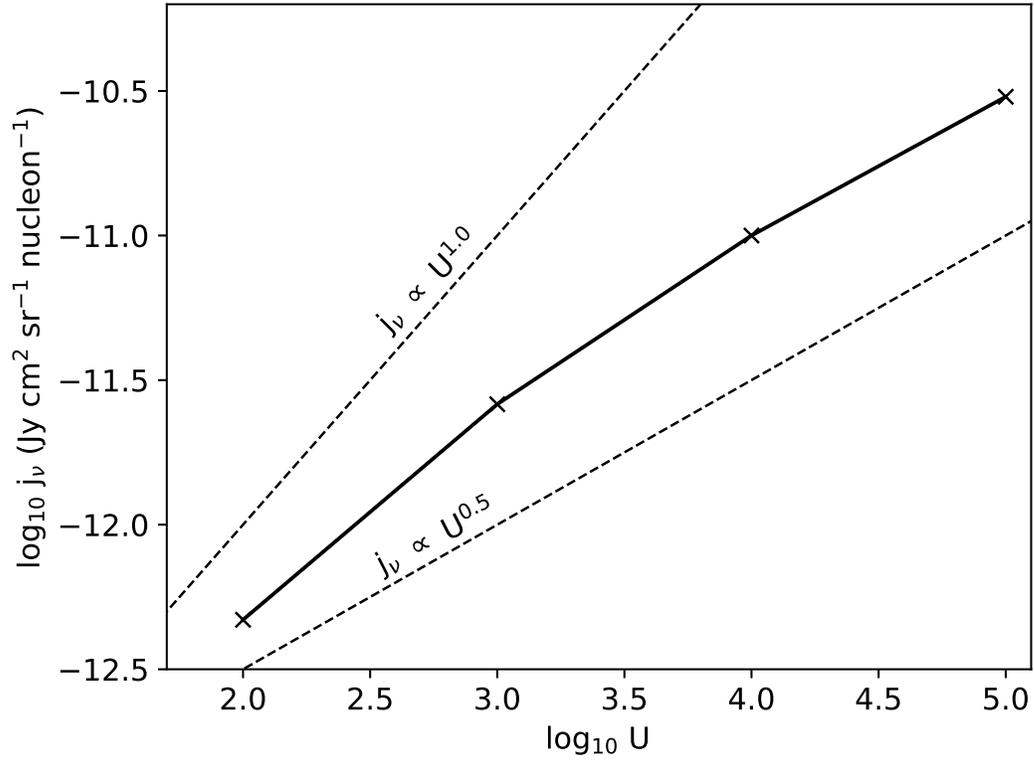}
\caption{Base-10 logarithm of the DL07 model emission coefficient, $j_\nu$,  
versus radiation density parameter, $U$ ({\it solid line and crosses}).  
The dashed lines show, for reference, power-law descriptions $j_\nu \propto U^{1.0}$ and 
$j_\nu \propto U^{1.0}$.
\label{fig:jnu}}
\end{figure}

Figure~\ref{fig:jnu} plots the
DL07 model emission coefficients versus the radiation
density parameter ({\it solid line and crosses}).   The
dashed lines show, for reference, power-law descriptions
$j_\nu \propto U^{1.0}$ and $j_\nu \propto U^{0.5}$.  
Figure~\ref{fig:jnu} shows that, over the range of $U$
covered by sample objects, $j_\nu$ is approximately
proportional to $\sqrt{U}$.  This means that $j_\nu$ is
relatively insensitive to the adopted $U$.  $U$ itself is
proportional to $R_*^2~T_{eff}^4/R_0^2$, where $R_*$ is the
stellar radius, $T_{eff}$ is the effective stellar
temperature, and $R_0$ is the standoff distance which we
previously expressed as $R_{0r} D$.  This means that our
estimate of $j_\nu$ implicitly contains a dependence on
these quantities,

\begin{equation}
j_\nu \propto \sqrt{U} \propto {\sqrt{{R_*^2~T_{eff}^4} \over {R^2_{0r} D^2}}} = {{R_*~T_{eff}^2} \over {R_{0r} D}}.
\end{equation}

\noindent It can now be seen that Equation~7 goes as,

\begin{equation}
 \dot M \propto { {{\bar m} {R^3_{0r}} D^2 V_a^2 I_\nu} \over {V_w \ell_{r}  R_* T_{eff}^2 }},
\end{equation}

\noindent so that this expression of the mass-loss rate
ultimately entails something close to a $D^2$ dependence, 
via the emission coefficient.  Accordingly, our analysis
here is restricted to objects that have well-constrained
distances.  If we characterize $j_\nu$ (via the DL07 models
and knowledge of $R_*$ and $T_{eff}$) as uncertain at the
30\% level,  the error budget for Equation 7 grows to about
70\%, or a factor of two if stellar velocity uncertainties
are included. Accordingly, we estimate the uncertainties on
$\dot M$ to be 0.3 dex for this sample.  Deviation  of the
mean stellar space velocities from the adopted $V_a$=30
\kms\  (not included in the above error budget) would
represent a systematic error shifting the  mass-loss rates
by a factor $V_{30}^2$, where $V_{30}$ is the relative
star-ISM velocity in units of 30 \kms. 

\section{Calculation of Mass-Loss Rates and Comparison to Prior Estimates}

Table~\ref{tab:derived} lists quantities derived from the 
basic data in Table~\ref{tab:basic}.  Columns 1--4 contain
the identifying numeral, name, generic name, and spectral
type, as in Table~\ref{tab:basic}.  Column~5 contains the
stellar luminosity in units of 10$^4$ solar luminosities.  
Column 6 contains the radiation density parameter, $U$,
calculated from the basic data.  Column 7 lists the 
corresponding emission coefficient interpolated from the 
DL07 models. Column 8 is the  ambient interstellar number
density, $n_a$, derived from the 70 $\mu$m specific
intensity, as described in the previous section.  Densities
range between 1.2 cm$^{-3}$ and 160 cm$^{-3}$, with a median
value of 16 cm$^{-3}$. These are typical of densities within
the cool neutral phase ($\approx$30 cm$^{-3}$) of the
interstellar medium and somewhat higher than the warm
neutral phase ($\approx$0.6 cm$^{-3}$) \citep[c.f.,][Table
1.3]{Draine2011}.   Column 9 contains the mass-loss rate
calculated from Equation 7.  Values range from 
2$\times$10$^{-9}$ \moy\ to 1.3$\times$10$^{-6}$ \moy, with
a median of 6$\times$10$^{-8}$ \moy. These are consistent
with the broad range of mass-loss rates for O stars found in
the literature and obtained using other methods.  

It is particularly instructive to compare our results  for
the well-studied prototypical bowshock star, $\zeta$ Oph
with other analyses.\footnote{We note here that the angular
standoff  distance for $\zeta$ Oph is incorrectly listed in
Table~5 \cite{Kobulnicky2017} and Table~1 of
\citet{Kobulnicky2016}  as 29\arcsec\ instead of
299\arcsec,  making the linear distance 0.159 pc for the
distance adopted here of 0.110 kpc.}   Our inferred density
of 2.3 cm$^{-3}$ compares  favorably with the 3.6 cm$^{-3}$
computed by \citet{Gvaramadze2012} and $\simeq$3 cm$^{-3}$
estimated from the radio free-free and H$\alpha$ surface
brightness of the surrounding \ion{H}{2} region
\citep{Gull1979}.  The resulting mass-loss rate is $\dot
M=$5.4$\times$10$^{-8}$ \moy. For comparison,
\citet{Gull1979} report $\dot M=$2.2$\times$10$^{-8}$ \moy\ 
using similar physical reasoning.   \citet{Gvaramadze2012}
list an identical $\dot M=$2.2$\times$10$^{-8}$ \moy.   Our
value is almost a factor of 30  larger than the $\dot
M=$1.8$\times$10$^{-9}$ \moy\ inferred by
\citet{Marcolino2009} by fitting model atmospheres to UV and
optical wind lines.  Our result is  a factor of 2.4 smaller
than the $\dot M=$1.3$\times$10$^{-7}$ \moy\ predicted from
the prescription of \citet{Vink2001} for the luminosity and
temperature of $\zeta$ Oph.  Predictions from the updated
moving reversing layer theory \citep{Lucy1970} by
\citet[][Table 1]{Lucy2010b}  for the adopted parameters of
$\zeta$ Oph ($T_{eff}=31,000$~K, $\log$ $g$=-3.75) indicate
$\dot M=$5.0$\times$10$^{-8}$ \moy, in excellent agreement
with our value.  This agreement is all the more impressive
given that $\zeta$ Oph is regarded as a weak-winded O star.
Given the general consistency of the momentum-balance
technique with theoretical expectations  and some other 
mass-loss measurements for this prototypical bowshock star,
we proceed to use the results from Table~\ref{tab:derived}
to assess its general applicability to mass loss from
massive stars.

Figure~\ref{fig:mdot} plots the calculated mass-loss rates
versus stellar effective temperature.\footnote{Although  the
mass-loss rate is expected to scale with {\it luminosity}
rather than  {\it temperature}, we choose here to plot the
latter to facilitate direct comparison with the
\citet{Lucy2010a} models and because most of our targets 
are of similar V--IV luminosity class.}    Black filled
symbols denote the 20 sample objects: a star for $\zeta$
Oph, circles for main-sequence stars,  and hexagons for
evolved stars.  Blue crosses depict model predictions for
each object using the formulation of \citet[][Equation
24]{Vink2001}  computed using the stellar data from Table~1
and assuming $v_{\infty}/v_{esc}$=2. 
Hence, each filled data point is paired
vertically with  a blue $x$ at the same temperature, although
the s's sometimes overlap.  Red squares  connected by lines
depict the model predictions from \citet{Lucy2010b} for
main-sequence ($\log$ $g$=4.0) and giant ($\log$ $g$=3.5)
stars, as labeled.  Red triangles and dotted lines  show the
predictions for B main-sequence stars from
\citet{Krticka2014}.   A cross above the legend depicts the
 typical measurement uncertainties on each
axis.  The dispersion in $\dot M$ at fixed temperature
is 0.35 dex, roughly consistent with our stated
uncertainties, but doubtless inflated  by the inclusion of a
six luminosity class IV/III objects among the sample of 20.
Key objects are labeled with spectral type and common
nomenclature.

Figure~\ref{fig:mdot} demonstrates that there is good 
agreement between the \citet{Vink2001} and \citet{Lucy2010b}
predictions.  Our new data fall $\approx$0.4 dex  below the
model predictions, but follow the same trend of increasing
mass-loss rate with effective temperature. The O4If star
BD+43 3654 lies two orders of magnitude below the  $\dot M >
10^5$ \moy\ levels predicted by the \citet{Lucy2010b}
relation for giants and an order of magnitude below the
\citet{Vink2001}  value for its temperature and
luminosity.     The O8V CPR2002 A10 and O7.5V/III BD+60 586
lie near the model  predictions but at the upper envelope of
the objects in our sample.  In the former case the spectral
classification comes from our own yellow--red optical
spectra \citep{Chick2018} which are not especially sensitive
to surface gravity.   This object could be of a more evolved
luminosity class which would lead to an expected mass-loss
rate more consistent with its position.    The {\it Spitzer
Space Telescope} 4.5/8.0/24 micron image of this object in
\citet{Kobulnicky2016}  reveals a very high surface
brightness nebula (indeed, the 70 $\mu$m  surface brightness
is the largest in our sample) that appears to be more like a
partial bubble  than a bowshock.  The inferred ambient
density of 72 cm$^{-3}$ is an outlier and is the second
largest in our sample.  If this is a windblown bubble,
meaning that the star's velocity actually quite low (perhaps
$<$10 \kms\ instead of the assumed 30 \kms) the resulting
mass-loss rate would drop by a factor of ten  into the
regime consistent with other stars of the O8V
classification. In the case of BD+60~586, \citet{Conti1974}
designate it as an O8III rather than O7.5V
\citep{Hillwig2006}, which would explain its position at the
high-$\dot M$ side of our sample. 

Objects having O9--O9.5 spectral types form a tightly
bunched vertical band in Figure~\ref{fig:mdot} covering the
range $10^{-8}$ \moy\ $< \dot M < 10^{-7}$ \moy, a factor of two--three
lower than  both sets of models, on average. The three
evolved stars (including $\zeta$ Oph)  lie toward the upper
end of this distribution. The dispersion in this subsample
is somewhat larger than the 0.3 dex  uncertainties,
suggesting some degree of variation in mass-loss rates in
this regime, although uncertainties on distance likely also
play a role given that $\dot M\propto D^2$.   We conclude
that the data for  late-O dwarfs (where the weak winds are
observed to be common and the  weak-wind problem is thought
to be most pronounced), subdwarfs, and one giant show nearly
an order of magnitude of dispersion and lie systematically
below model predictions by a factor of about two, on average.

For the three early-B stars the situation is less clear. The
B0IVe star HD53367 lies in the lower left of
Figure~\ref{fig:mdot} at $\dot M=$2.2$\times$10$^{-9}$ \moy,
well below two of the three models, but  in excellent
agreement with the \citet{Krticka2014} prediction.   The
B0III star FN CMa lies nearly an order of magnitude above
model \citet{Lucy2010b} expectations and above the other
data points but quite near the \citet{Vink2001} prediction
at  $\dot M=$2.9$\times$10$^{-7}$ \moy.  Its prominent
bowshock nebula appears well-defined and
well-characterized.  Its distance is somewhat poorly
constrained by parallax at 0.94$^{+1.1}_{-0.47}$ kpc.   This
may be a case where the space velocity is significantly
different than the assumed 30 \kms.  If its space velocity
were to be much lower, perhaps 10 \kms, this object would be
consist with \citet{Lucy2010b} model expectations and with
extrapolation of the trend defined by late-O stars. 
Finally, the B1V star KGK2010 2 lies an order of magnitude
above the \citet{Vink2001} prediction, outside of the regime
of the \citet{Lucy2010b} models,  and two orders of
magnitude above the \citet{Krticka2017} prediction.   This
nebula has the third highest surface brightness in our
sample and has one of the smallest standoff distance
distances, making it very compact.  Our multiple optical
spectra of this star allow a range of spectral types,
B2--B0, but the luminosity class is not well constrained. 
Its reddening and broadband magnitudes make it consistent
with an early-B star at the 1.32 kpc distance of Cygnus
OB2.  A larger distance would only exacerbate the extremity
of its apparent mass loss rate.  The infrared images of its
nebula in \citet{Kobulnicky2016, Kobulnicky2017} show a
strikingly bright and bowshock-like morphology visible at
3.6 $\mu$m through 160 $\mu$m, making it one of the few
objects among the 709-object bowshock catalog
\citep{Kobulnicky2016}  detected across all seven $SST$ and
$HSO$ infrared bandpasses.  The inferred ambient number
density of  160 cm$^{-3}$  is, by far, the largest in our
sample. This, coupled with the detection at even the
shortest  $SST$ bandpasses suggests an unusual interstellar
environment.  This object may be running into a molecular
cloud, for instance.  \citet{Kobulnicky2016} note that  the
infrared SED is one of the few objects better fit by a dust
model with large PAH fraction, $q_{PAH}$=4.58\%.  It has the
coolest 24-to-70 $\mu$m color temperatures among the sample
\citep[$T_{24/70}$=70~K, c.f.,][Table 5]{Kobulnicky2017}. 
We attempted using DL07 models having a larger PAH fraction
with similar radiation density, but these yield smaller
emission coefficients, which only serve to increase the
resulting mass-loss rate.   We conclude that the DL07 dust
models may not be adequate in this case.  Perhaps the PAHs
and dust at the surface of a colder molecular structure are
being fragmented in the wind shock so that the interstellar
dust grain size distribution or grain composition  adopted
by DL07 is not appropriate here.

Figure~\ref{fig:mdot2} replicates Figure~\ref{fig:mdot} with
the addition mass loss rates measured for the set of
galactic O3--O9.5 dwarf stars studied separately by
\citet{Martins2005b} ({\it blue open circles})  and
\citet{Howarth1989} ({\it blue open squares}),
respectively.  The discrepancy discussed in 
\citet{Martins2005b} is obvious here, with the open squares
lying 0.5--1.5 dex above the open circles.\footnote{We have
shifted the effective temperatures  assigned by
\citet{Martins2005b} by $-$2000 K for consistency with the
O9.5V objects in our sample.}   The \citet{Martins2005b}
mass-loss rates derived from UV spectra appear consistent
with the bowshock sample at the upper end of the temperature
range but lie well below the bowshock sample in the O8--O9
regime.   The mass-loss rates given by  \citet{Howarth1989} 
are consistently higher than the observational results
presented here at the same spectral type ({\it black filled
symbols}), but there is considerable scatter and some
overlap. The \citet{Howarth1989} values are broadly
consistent with the   theoretical expectations for dwarfs.  
Figure~\ref{fig:mdot2} also shows the late-O dwarf and giant
stars  with mass-loss rates determined from ultraviolet
P$^{4+}$ absorption lines \citep[][{\it green x's and +'s,
respectively}]{Fullerton2006}  and the same set of stars
determined from the H$\alpha$ line \citep[][{\it cyan x's
and +'s}]{Repolust2004,Markova2004}. The P$^{4+}$
measurements show a large dispersion at any given effective
temperature, but generally lie an order of magnitude below
the bowshock sample.   The H$\alpha$ results lie
significantly above the bulk of the data and models,
although most of the points for dwarfs are upper limits so,
thereby, formally consistent with the other data without
providing strong constraints.  These upper limits underscore
the  difficulty in measuring mass-loss rates for late-O
stars using the H$\alpha$  line.  

\section{Discussion and Conclusions} 

\begin{figure}
\plotone{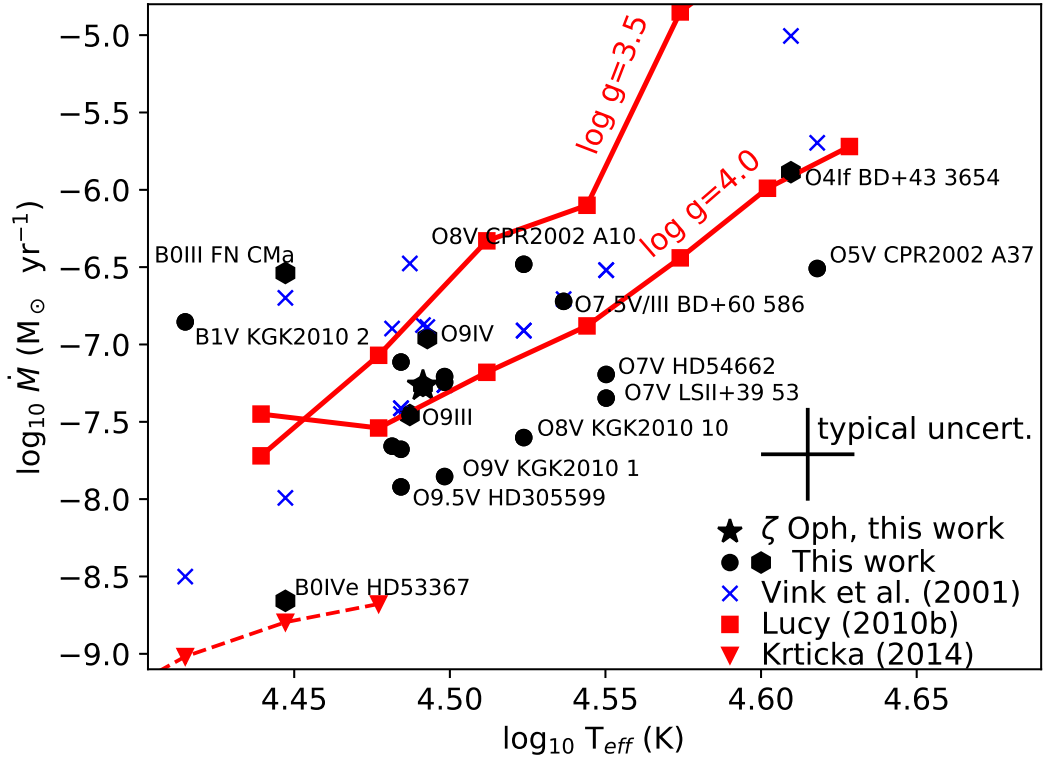}
\caption{Mass-loss rate versus stellar effective temperature.  
Black filled points and hexagons depict measurements for main-sequence and evolved 
stars from this sample, respectively.  Blue crosses mark predictions for each object using the 
expression of \citet[][Equation 24]{Vink2001}.  Red squares and lines show the model predictions of \citet{Lucy2010b}
for nominal main-sequence ($\log$ g=4.0) and giant ($\log$ g=3.5) stars, as labeled. The triangles and
dotted lines show the theoretical predictions of \citet{Krticka2014} for B stars.   
\label{fig:mdot}}
\end{figure}

Mass-loss rates derived from the principle of momentum
balance and those predicted by two theoretical frameworks
\citep{Vink2001, Lucy2010b} in Figure~\ref{fig:mdot} display
a similar trend with effective temperature but are offset by
about an average factor of about two lower.  Knowledge of
the star's velocity, stellar wind velocity, ambient density,
and bowshock size, yield mass-loss rates in good agreement
with the \citet{Howarth1989} UV analysis, but substantially
larger than more modern analyses of UV spectra in
conjunction with atmosphere models such as CMFGEN
\citep{Hillier1998}.  Our results are factors of several lower than
H$\alpha$-based measurements, uncorrected for clumping, consistent with current 
consensus that a correction by factors of several for clumping is required.  
 That the dispersion in present results
and  the overall slope and zero point  of the $\dot M$-$T_{eff}$ relation is
similar to other techniques and models suggests promise for 
the momentum-balance method, employed here
for the first time using a sizable sample.  Concomitantly,
this could be seen as an  indirect validation of the DL07
emission coefficients for dust  within bowshock nebulae---an
environment where  it cannot be taken for granted that the
prescriptions for typical interstellar dust size
distributions, compositions, PAH absorption cross sections,
grain heat capacities, dielectric functions, etc., will
apply.  It would not be unreasonable, {\it a priori}, to
expect that within bowshock nebulae shocks act to fragment
grains (as we speculate in the case of KGK2010 2) so that
the DL07  models are inappropriate.  This is evidently not
the case for the majority of our sample.  Viewed from
another perspective, the general agreement in
Figure~\ref{fig:mdot} could be seen as a validation of the
\citep{Vink2001} and \citet{Lucy2010b}  theoretical
predictions using an observational technique that is
unaffected by effects like clumping that plague
density-squared diagnostics or the ionization structure
uncertainties associated with absorption line diagnostics. 
It remains unclear whether the 0.2--0.4 dex offset between
the theoretical expectations and  the bowshock sample can
best be reconciled by identifying a systematic problem with
the bowshock formalism outlined here or by further
refinement in the theoretical treatment of stellar winds.  
Mass-loss rates derived here lie closer to model
expectations than other observational results, ameliorating,
but not fully resolving, the weak wind problem.   This
qualified success of the momentum-balance approach  may be
used to refine traditional mass loss diagnostics for 
application to stars which lack bowshock nebulae (the vast
majority!).

\begin{figure}
\plotone{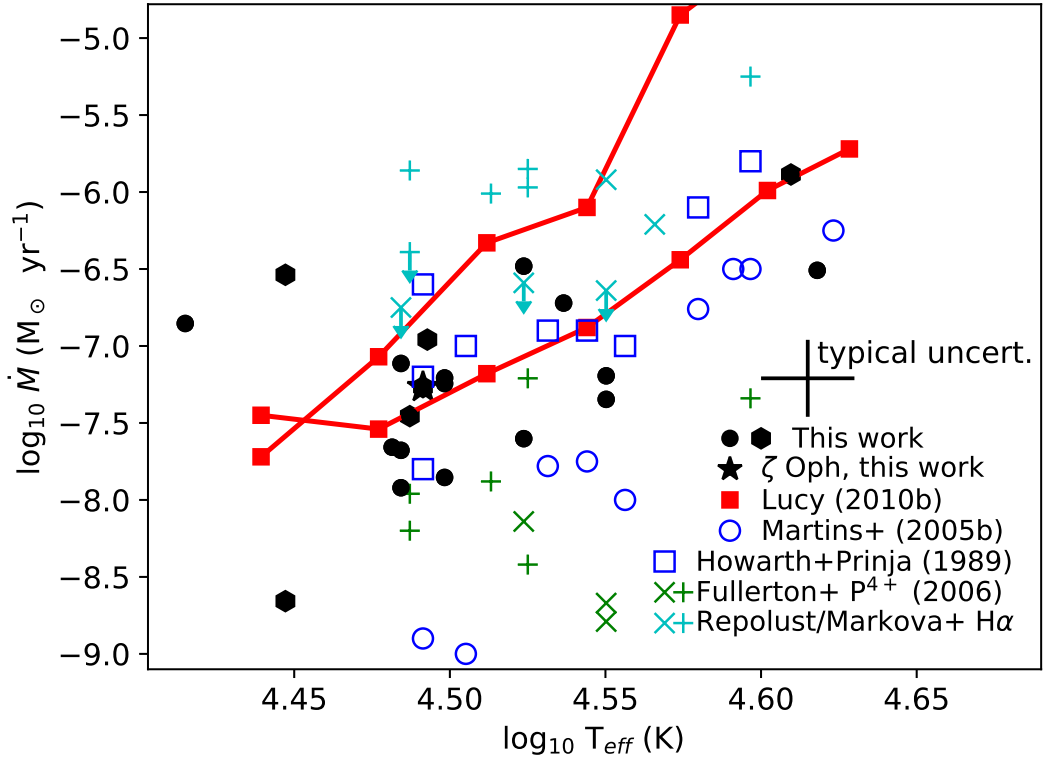}
\caption{Mass-loss rate versus stellar effective temperature, with symbols
as in Figure~\ref{fig:mdot}.  
Open circles and squares depict the sample of galactic O3--O9 main-sequence stars as
measured by \citet{Martins2005b} and \citet{Howarth1989}, respectively. 
Green x's and +'s depict the dwarfs and giants, respectively, measured using the ultraviolet P$^{4+}$ line  
\citep{Fullerton2006}.  Cyan x's and +'s depict the same stars as determined from the H$\alpha$ line.
\label{fig:mdot2}}
\end{figure}

The disparity between various $\dot M$ determinations 
evidenced in Figure~\ref{fig:mdot2} reflects the historical
measurement problems discussed extensively in the literature
regarding late-O stars.  Resolving disparities with $n^2$
methods such as H$\alpha$ is generally attempted by invoking
{\it ad hoc} wind clumping factors of several
\citep{Fullerton2006,Prinja2010}.  Resolving  disparities
with UV absorption diagnostics coupled with theoretical model
atmospheres has been attempted by invoking modifications of
the ionization structure \citep{Lucy2010b} or X-ray ionized
winds \citep{Huenemoerder2012}.   While the new measurements
here do not help identify specific problems with classical $\dot
M$ measurements, the self-consistency, lack of adjustable
parameters, and better agreement with  recent theoretical
developments may help in revising those techniques.  

The inferred interstellar ambient densities in
Table~\ref{tab:derived} provide some insight regarding the
conditions where bowshocks form.   \citet{Peri2012} and
\citet{Peri2015} concluded that only 10--15\% of 
high-proper motion massive stars showed evidence of infrared
bowshocks.  \citet{Huthoff2002} argued that  that
interstellar density  likely plays a larger role than space
velocity or stellar spectral type in creating  an observable
nebula, a conclusion supported by hydrodynamical simulations
of \citet{Comeron1998} and \citet{Meyer2016}. They found
that slightly supersonic velocities, strong stellar winds, 
and larger ambient  interstellar densities
$n_a>$0.1~cm$^{-3}$  resulted in the  visible bowshock
nebulae.  Our range of densities runs from $n_a=$1.2 to 160
cm$^{-3}$ with a mean of $n_a=$24 cm$^{-3}$.  $\zeta$ Oph
has the second lowest density with $n_a=$2.3 cm$^{-3}$.  
Its 70 $\mu$m surface brightness is the lowest in our
sample, leading \citet{Kobulnicky2017} to note that it would
not likely be detectable if it were not located 23\degr\
above the Galactic Plane in a region of low infrared
background.  For most of the objects which lie within a
degree of the Plane, we infer that ambient densities of at
least $n_a\gtrsim$5 cm$^{-3}$ appear to be required. 

The uncertainties on our measurements are dominated, at
present, by the lack of data on space velocities for each
star in the frame of the local interstellar medium.  Because
mass-loss rates scale as $V_a^2$, our adoption of a single
$V_a$=30 \kms\ likely leads to significant errors in some
cases.  True three-dimensional space velocities and accurate
distances, such as will be provided by the GAIA mission
\citep{Gaia2016a}, should be available for many stars of
interest in the near future so that more precise mass loss
rates will be possible.  We are presently conducting a
spectroscopic survey of bowshock stars that will provide
needed data (spectral classifications, stellar temperatures,
radii, radial velocities) for a much larger sample.   An
enhanced  sample of early-B stars will result from this work
so that  mass-loss rates in this low-temperature, weak-wind
regime should finally be possible.  

\acknowledgments This work has been supported by the
National Science Foundation through grants AST-1412845 and 
AST-1560461 (REU).  We thank Nathan Smith and an anonymous
reviewer for suggestions that improved 
this manuscript.

\vspace{5mm}
\facilities{SST, WISE, HSO, HIPPARCOS}

\newpage

\begin{deluxetable}{rcrcrrrrrrrrrrr}
\tablecaption{Measured \& adopted parameters for stars and their bowshock nebulae \label{tab:basic}}
\rotate
\tablehead{
\colhead{ID} &\colhead{Name}&\colhead{Alt. name}&\colhead{Sp.T.}&\colhead{Ref.}&\colhead{T$_{eff}$}&\colhead{R$_*$ }	 &\colhead{Mass}&\colhead{$V_\infty$}  &\colhead{D}   &\colhead{Ref. }&\colhead{$R_0$}	&\colhead{$R_0$}&\colhead{Peak$_{70}$}  	&\colhead{$\ell$}   \\
\colhead{}   &\colhead{    }&\colhead{         }&\colhead{     }&\colhead{ }   &\colhead{(K)}	   &\colhead{($R_\odot$)}&\colhead{(M$_\odot$)}    &\colhead{(\kms)}       &\colhead{(kpc)}&\colhead{ }  &\colhead{(arcsec)} &\colhead{(pc)} &\colhead{(10$^7$ Jy sr$^{-1}$)}&\colhead{(arcsec)} \\
\colhead{(1)}&\colhead{(2)} &\colhead{(3)}	&\colhead{(4)}  &\colhead{(5)} &\colhead{(6)}	   &\colhead{(7)}	 &\colhead{(8)}       &\colhead{(9)}	      &\colhead{(10)}  &\colhead{(11)}&\colhead{(12)}	 &\colhead{(13)} &\colhead{(14)}		&\colhead{(15)}   
}
\startdata
 13 & $\zeta$ Oph    & G006.2812+23.5877   & O9.2IV                    & S1   & 31000 & 10  & 19  & 1300 & 0.11 & D1		   &299 & 0.175 & 12.3 & 277 \\
 67 & NGC 6611 ESL 45& G017.0826+00.9744   & O9V                       & S2   & 31500 & 7.7 & 18  & 1300 & 1.99 & D2		   &7.5 & 0.080 & 64.0 & 15 \\
329 & KGK 2010 10    & G077.0505$-$00.6094 & O8V                       & S3   & 33400 & 8.5 & 23  & 2000 &  4.1 & D3\tablenotemark{a}&10  & 0.219 & 15.4 & 27 \\
331 & LS II +39 53   & G078.2869+00.7780   & O7V                       & S4   & 35500 & 9.3 & 26  & 2500 & 1.32 & D3\tablenotemark{a}&42  & 0.296 & 12.0 & 55 \\
338 & CPR2002A10     & G078.8223+00.0959   & O8V:                      & S3   & 33400 & 8.5 & 23  & 1200 & 1.32 & D3\tablenotemark{a}&23  & 0.162 & 79.8 & 29 \\
339 & CPR2002A37     & G080.2400+00.1354   & O5V((f))                  & S5   & 41500 & 11.1& 37  & 2900 & 1.32 & D3\tablenotemark{a}&70  & 0.493 & 29.9 & 47 \\
341 & KGK2010 1      & G080.8621+00.9749   & O9V                       & S3   & 31500 & 7.7 & 18  & 1300 & 1.32 & D3\tablenotemark{a}&20  & 0.141 & 5.0  & 31 \\
342 & KGK2010 2      & G080.9020+00.9828   & B1V:                      & S3   & 26000 & 6   & 10  &  800 & 1.32 & D3\tablenotemark{a}&10  & 0.070 & 57.8 & 14 \\
344 & BD +43 3654    & G082.4100+02.3254   & O4If                      & S6   & 40700 & 19  & 58  & 3000 & 1.32 & D3\tablenotemark{a}&193 & 1.359 & 58.6 & 170 \\
368 & KM Cas         & G134.3552+00.8182   & O9.5V((f))                & S7   & 30500 & 7.4 & 16  & 1200 & 1.95 & D4\tablenotemark{b}&14  & 0.146 & 24.9 & 22 \\
369 & BD +60 586     & G137.4203+01.2792   & O7.5V/O8III               & S8   & 34400 & 8.9 & 24  & 2500 & 1.95 & D4\tablenotemark{b}&73  & 0.759 & 7.9  & 39 \\
380 & HD 53367       & G223.7092$-$01.9008 & B0IVe                     & S9   & 28000 & 7   & 15  & 1200 & 0.26 & D1		   &15  & 0.021 & 35.7 & 49 \\
381 & HD 54662       & G224.1685$-$00.7784 & O7Vzvar?\tablenotemark{c} & S1   & 35500 & 9.4 & 26  & 2500 & 0.63 & D1		   &220 & 0.739 & 4.6  & 200\\
382 & FN CMa         & G224.7096$-$01.7938 & B0III                     & S9   & 28000 & 15  & 20  & 1200 & 0.94 & D1		   &101 & 0.506 & 11.2 & 70 \\
406 & HD 92607       & G287.1148$-$01.0236 & O9IVn                     & S1   & 31100 & 10  & 20  & 1300 & 2.35 & D5\tablenotemark{d}&16  & 0.201 & 29.1 & 26 \\ 
407 & HD 93249       & G287.4071$-$00.3593 & O9III+O:                  & S1   & 30700 & 13.6& 23  & 1300 & 2.35 & D5\tablenotemark{d}&7.8 & 0.098 & 58.2 & 25 \\
409 & HD 93027       & G287.6131$-$01.1302 & O9.5IVvar\tablenotemark{e}& S10  & 30300 & 10  & 16  & 1200 & 2.35 & D5\tablenotemark{d}&7.4 & 0.093 & 20.8 & 17 \\
410 & HD 305536      & G287.6736$-$01.0093 & O9.5V+?\tablenotemark{f}  & S1   & 30500 & 7.4 & 15  & 1200 & 2.35 & D5\tablenotemark{d}&3.7 & 0.046 & 91.4 & 14 \\
411 & HD 305599      & G288.1505$-$00.5059 & O9.5V                     & S11  & 30500 & 7.4 & 15  & 1200 & 2.35 & D5\tablenotemark{d}&4.2 & 0.052 & 41.5 & 16 \\
413 & HD 93683       & G288.3138$-$01.3085 & O9V+B0V\tablenotemark{g}  & S11  & 31500 & 7.7 & 18  & 1300 & 2.35 & D5\tablenotemark{d}&15  & 0.188 & 15.4 & 24 \\
\enddata
\tablecomments{(1) Identifier from \citet{Kobulnicky2016}, (2) Common name, (3) generic identifier in galactic coordinates, (4) spectral classification, (5) reference for
spectral classification, (6) effective temperature based on spectral classification using the theoretical scale of \citet{Martins2005a}, (7) stellar radius
 based on spectral classification using the theoretical scale of \citet{Martins2005a}, (8) adopted stellar mass, (9) adopted terminal wind velocity from \citet{Mokiem2007}, (10) adopted distance, 
 (11) reference for distance, (12) standoff distance in arcsec, (13) standoff distance in pc
 using the adopted distance and angular size from \citet{Kobulnicky2017} adjusted by a statistical factor of 1.1 for projection effects, (14) peak 70 $\mu$m surface 
 brightness above adjacent background, (15) angular diameter of the nebula in arcsec defined by a chord intersecting the location of peak surface brightness.
References for spectral types: S1--\citet{Sota2014}; S2--\citet{Evans2005}; S3--\citet{Chick2018}; S4--\citet{Vijapurkar1993}; S5--\citet{Hanson2003}; 
  S6--\citet{Comeron2007}; S7--\citet{Massey1995}; S8--\citet{Hillwig2006} gives O7.5V but \citet{Conti1974} lists O8III; S9--\citet{Tjin2001}, S10--\citet{Sota2011}, S11--\citet{Alexander2016}
  References for Distances: D1--\citet{Vanleeuwen2007} ; D2---\citet{Hillenbrand1993}; D3--\citet{Kiminki2015}; D4--\citet{Xu2006}, D5--\citet{Smith2006} 
   }
\tablenotetext{a}{Assumed to be at a similar distance as Cygnus OB2 \citep{Kiminki2015} based on similar magnitude and reddening, but see notes in text on KGK2010 10.}
\tablenotetext{b}{Assumed to be in the Perseus spiral arm as part of the Cas OB6 Association near the W3/W4/W5 star forming regions 
  having maser parallax measurements by \citet{Xu2006}. This is consistent with the the open cluster photometric distance of 2.2$\pm$0.2 kpc \citep{Lim2014}.}
\tablenotetext{c}{The possible double-lined nature (O6.5V+O7-9V 2119 d period)  of this source reported by \citet{Boyajian2007} but was not confirmed by \cite{Sota2014}.}
\tablenotetext{d}{Understood to be part of the Carina Nebula complex at 2.35 kpc distance \citet{Smith2006}, consistent with other contemporary determinations.}
\tablenotetext{e}{A single-lined eclipsing binary according to \citet{Sota2011}, suggesting a significant difference in mass between the primary and secondary star. }
\tablenotetext{f}{A possible single-lined spectroscopic binary according to \citet{Levato1990}. }
\tablenotetext{f}{A double-lined spectroscopic binary according to \citet{Alexander2016}. }
\end{deluxetable}

\newpage

\begin{deluxetable}{rcrcrrrrrr}
\tablecaption{Derived parameters for stars and their bowshock nebulae \label{tab:derived}}
\tablehead{
\colhead{ID} &\colhead{Name}&\colhead{Alt. name}&\colhead{Sp.T.}&\colhead{Lum.}              &\colhead{$U$}&\colhead{$j_\nu$}             &\colhead{$n_a$}      &\colhead{$\dot M$}		\\
\colhead{}   &\colhead{    }&\colhead{         }&\colhead{     }&\colhead{(10$^4$ L$_\odot$)}&\colhead{}   &\colhead{(Jy sr $^{-1}$ cm$^2$ nuc$^{-1}$)} &\colhead{(cm$^{-3}$)}&\colhead{(M$_\odot$~yr$^{-1}$)} \\
\colhead{(1)}&\colhead{(2)} &\colhead{(3)}	&\colhead{(4)}  &\colhead{(5)}               &\colhead{(6)}&\colhead{(7)}                  &\colhead{(8)}        &\colhead{(9)}		      
}
\startdata 
 13 & $\zeta$ Oph    & G006.2812+23.5877   & O9.2IV          & 8.1 & 4.2$\times10^3$ & 8.7$\times10^{-12}$ & 2.3& 5.4$\times$10$^{-8}$ \\
 67 & NGC 6611 ESL 45& G017.0826+00.9744   & O9V             & 5.1 & 1.3$\times10^4$ & 1.1$\times10^{-11}$ & 32 & 6.2$\times$10$^{-8}$  \\
329 & KGK 2010 10    & G077.0505$-$00.6094 & O8V+?           & 7.9 & 2.7$\times10^3$ & 8.4$\times10^{-12}$ & 1.3& 2.5$\times$10$^{-8}$  \\
331 & LS II +39 53   & G078.2869+00.7780   & O7V             & 12.0& 2.2$\times10^3$ & 8.3$\times10^{-12}$ & 6.6& 4.5$\times$10$^{-8}$  \\
338 & CPR2002A10     & G078.8223+00.0959   & O8V:            & 7.9 & 4.9$\times10^3$ & 8.9$\times10^{-12}$ & 72 & 3.3$\times$10$^{-7}$  \\
339 & CPR2002A37     & G080.2400+00.1354   & O5V((f))        & 32.0& 2.1$\times10^3$ & 8.3$\times10^{-12}$ & 21 & 3.1$\times$10$^{-7}$  \\
341 & KGK2010 1      & G080.8621+00.9749   & O9V             & 5.1 & 4.2$\times10^3$ & 8.7$\times10^{-12}$ & 15 & 1.4$\times$10$^{-8}$  \\
342 & KGK2010 2      & G080.9020+00.9828   & B1V:            & 1.4 & 4.7$\times10^3$ & 8.9$\times10^{-12}$ & 160& 1.4$\times$10$^{-7}$  \\
344 & BD +43 3654    & G082.4100+02.3254   & O4If            & 87.0& 7.6$\times10^2$ & 8.0$\times10^{-12}$ & 19 & 1.3$\times$10$^{-6}$  \\
368 & KM Cas         & G134.3552+00.8182   & O9.5V((f))      & 4.1 & 3.2$\times10^3$ & 8.5$\times10^{-12}$ & 21 & 7.7$\times$10$^{-8}$  \\
369 & BD +60 586     & G137.4203+01.2792   & O7.5V           & 9.7 & 2.7$\times10^2$ & 7.8$\times10^{-12}$ & 18 & 1.9$\times$10$^{-7}$  \\
380 & HD 53367       & G223.7092$-$01.9008 & B0IVe           & 2.6 & 9.9$\times10^4$ & 3.0$\times10^{-11}$ & 16 & 2.2$\times$10$^{-9}$  \\
381 & HD 54662       & G224.1685$-$00.7784 & O7Vzvar?        & 12.0& 3.7$\times10^2$ & 7.9$\times10^{-12}$ & 1.2& 6.4$\times$10$^{-8}$  \\
382 & FN CMa         & G224.7096$-$01.7938 & B0III           & 12.0& 7.7$\times10^2$ & 8.0$\times10^{-12}$ & 18 & 2.9$\times$10$^{-7}$  \\
406 & HD 92607       & G287.1148$-$01.0236 & O9IVn           & 8.2 & 3.3$\times10^3$ & 8.5$\times10^{-12}$ & 16 & 1.1$\times$10$^{-7}$  \\ 
407 & HD 93249       & G287.4071$-$00.3593 & O9III+O:        & 14.0& 2.5$\times10^4$ & 1.3$\times10^{-11}$ & 12 & 3.5$\times$10$^{-8}$  \\
409 & HD 93027       & G287.6131$-$01.1302 & O9.5IVvar       & 7.4 & 1.4$\times10^4$ & 1.1$\times10^{-11}$ & 9  & 2.2$\times$10$^{-8}$  \\
410 & HD 305536      & G287.6736$-$01.0093 & O9.5V+?         & 4.1 & 3.1$\times10^4$ & 1.5$\times10^{-11}$ & 29 & 2.1$\times$10$^{-8}$  \\
411 & HD 305599      & G288.1505$-$00.5059 & O9.5V           & 4.1 & 2.4$\times10^4$ & 1.3$\times10^{-11}$ & 6  & 1.2$\times$10$^{-8}$  \\
413 & HD 93683       & G288.3138$-$01.3085 & O9V+B0V         & 5.1 & 2.4$\times10^3$ & 8.3$\times10^{-12}$ & 11 & 5.7$\times$10$^{-8}$  \\
\enddata
\tablecomments{(1) Identifier from \citet{Kobulnicky2016}, (2) Common name, (3) generic identifier in galactic coordinates, (4) spectral classification, (5)
stellar luminosity computed from effective temperature and radius in Table~\ref{tab:basic}, (6) dimensionless ratio of the radiant energy
density (in erg~cm$^{-3}$) from to the star to the mean
interstellar radiant energy density estimated by \citet[][MMP83]{Mathis1983}, as tabulated by \citet{Kobulnicky2017}, (7)   dust emission coefficient 
 expressing the  energy emitted per second per steradian per Hertz per {\it nucleon}, (8) ambient interstellar number density, computed from Equation 5,
 (9) computed mass-loss rate in solar masses per year.} 
\end{deluxetable}

\newpage

\newpage

\newpage

\newpage

\newpage

\end{document}